 \documentstyle[11pt,doublespace]{article}

\begin{document}


\rightline{LA-UR-95-241}

\begin{center}

{\bf\large {Quantum Computers, Factoring and Decoherence}}
\vspace*{1.1ex}

{I. L. Chuang$^1$, R. Laflamme$^2$, P. Shor$^3$
			\/{\em and}\/ W. H. Zurek$^2$}\\

{\vspace*{1.0ex}
	$^1$ Edward L. Ginzton Laboratory, Stanford University,
	Stanford, CA, 94305, USA	\\[1.2ex]
	$^2$Theoretical Astrophysics, T-6, MS B288	\\
	Los Alamos National Laboratory, Los Alamos, NM87545, USA \\[1.2ex]
	$^3$AT\& T Bell Labs, 600 Mountain Ave., Murray Hill, NJ 07974, USA
\\[1.2ex]
}
{February 5, 1995}

\end{center}



\begin{singlespace}

\begin{center}
{\bf Abstract}
\end{center}

\begin{quote}
\small

In a quantum computer any superposition of inputs evolves unitarily
into the corresponding superposition of outputs.  It has been recently
demonstrated that such computers can dramatically speed up the task of
finding factors of large numbers -- a problem of great practical
significance because of its cryptographic applications.  Instead of
the nearly exponential ($\sim \exp L^{1/3}$, for a number with $L$
digits) time required by the fastest classical algorithm, the quantum
algorithm gives factors in a time polynomial in $L$ ($\sim L^2$).
This enormous speed-up is possible in principle because quantum
computation can simultaneously follow all of the paths corresponding
to the distinct classical inputs, obtaining the solution as a result
of coherent quantum interference between the alternatives.  Hence, a
quantum computer is sophisticated interference device, and it is
essential for its quantum state to remain coherent in the course of
the operation.  In this report we investigate the effect of
decoherence on the quantum factorization algorithm and establish an
upper bound on a ``quantum factorizable'' $L$ based on the decoherence
suffered per operational
step.

\end{quote}
\end{singlespace}


The uniqueness of the prime factorization of a positive integer is the
Fundamental Theorem of Arithmetic\cite{factoring}.  However, in
practice, the determination of the prime factors of a given number can
be an exceedingly difficult problem, although verification is trivial.
This asymmetry is the basis for modern cryptography, and provides
secret codes used not only on your own bank card but also to transfer
diplomatic messages between embassies.

Attempts to undermine the security provided by the difficulty of
factorization have met with failure by and large, even with the aid of
powerful modern computers.  In fact, this problem is widely believed
to have no polynomial-time algorithm\cite{npprob}, although a proof of
this statement has remained elusive.  The best known classical
computer algorithm\cite{Lenstra93} to factor a number $N$ of $L$
digits takes a time exponential in $L^{1/3}$.

In contrast, one of us\cite{Shor94} has shown recently that with the
help of a quantum computer one can factor numbers in a random {\em
polynomial\/} amount of time.  Therefore these new computers could be
a threat to what is presently the most common method of encrypted
message transfer.  However, it is still unknown whether such machines
are {\em practical}, because they depend crucially on
quantum-mechanical behavior which is uncommon to our mostly classical
world.  This issue is one of decoherence\cite{Zurek91},
the subject of our study.

The quantum factoring algorithm uses in an essential way the coherence
of a quantum wavefunction.  In a nutshell, to factor a number $N$ one
chooses a number $x$ at random and calculates its order $r$ modulo
$N$, i.e.  finds $r$ such that $x^r \equiv\ 1\ {\rm mod}\ N$. Once $r$
is known, factors of $N$ may often be found using the Chinese
remainder theorem.  The difficulty is to calculate $r$.  The quantum
factoring algorithm goes as follows.  First choose a smooth number
(one with small prime factors) $q$ such that $N^2<q<2N^2$ and build
the state;
\begin{equation}
	|\psi_1\rangle = {1\over\sqrt{q}} \sum_{a=0}^{q-1} |a,0\rangle ,
\end{equation}
from which can be obtained (using a quantum computer)
\begin{equation}
	|\psi_2\rangle = {1\over \sqrt{q}}
			\sum_{a=0}^{q-1} |a,x^a\ {\rm mod}\ N\rangle.
\end{equation}
We can now Fourier transform this pure state (again using a quantum
computer) to get;
\begin{equation}
	| \psi_3\rangle = {1\over q} \sum_{c=0}^{q-1}\sum_{a=0}^{q-1}
                e^{i2\pi ac/q} |c,x^a\ {\rm mod}\ N\rangle ,
\label{eq:shorstate}
\end{equation}
and measure both arguments of this superposition, obtaining $\bar c$
for the first one and some $x^k$ as the answer for the second one ($k$
being any number between 0 and $r$).  Given the pure state
$|\psi_3\rangle$, probabilities of different results for this
measurement will be given by the probability distribution;
\begin{equation}
	P({\bar c},x^k) = \left| {1\over q}
		\sum_{a=0}^{q-1}
		{\mbox{\hspace*{-2pt}\raisebox{3pt}{$\mathop{'}$}}}
			e^{i2\pi a{\bar c}/q}
			\right|^2
\,,
%
%
\end{equation}
where the prime indicates a restricted sum over values of $a$ which
satisfy $x^a \equiv x^k {\rm mod}\ N$.  This function has periodicity
$q/r$, but as we know $q$, we can determine $r$ with a few trial
executions (an example is shown in Fig.~\ref{fig:perfect}).  A
measurement thus gives with high probability $c=\lambda q/r$, where
$\lambda$ is an integer which corresponds to a particular peak in
Fig.~\ref{fig:perfect}.  With a few runs of the program, we can deduce
$r$ and thus the factors of $N$.

The algorithm discussed above assumes that the quantum computer was
completely isolated.  In practice this will certainly not be the case.
It is the effect of imperfect isolation which we study here.  A first
obvious effect is that the quantum computer will lose energy.  This
happens at the rate $\tau_{rel}$, the relaxation time-scale.  It is
relatively easy to make systems for which $\tau_{rel}$ can be very
large and thus allow a reasonable number of operation to complete.  A
much more insidious effect of imperfect isolation is {\em
decoherence}[5].  Decoherence is caused by the continuous interaction
between the system (in our case the quantum computer) and the
environment[5-7].  As a result, the state of the
environment ``monitors,'' and therefore becomes correlated with, the
state of the system.  As a quantum system evolves, information about
its states leaks out into the environment, causing them to loose their
purity, and, consequently, their ability to interfere.

It is important to realize that the timescale for decoherence
$\tau_{dec}$ is much smaller than the one for relaxation.  For
example, an oscillator of mass $m$ in a superposition of coherent
states (separated by a distance $\Delta x$ from each other)
interacting linearly with a bath at temperature $T$ has the
decoherence time\cite{Zurek86b}
\begin{equation}
	\tau_{dec} \sim \tau_{rel}
		\left [ {\lambda_{dB} \over \Delta x } \right ]^2
\,,
\end{equation}
where $\lambda_{dB}$ is the thermal de Broglie wavelength.  This expression is
valid for high temperatures only; at low temperatures, the
$\tau_{dec}$ becomes inversely proportional to the cut-off frequency
of the bath.  It is crucial to realize that no net energy transfer is
needed to effect decoherence.  This implies a much greater sensitivity
of quantum computation to decoherence than to the relaxation process.

The decoherence process has been proposed as a mechanism for enforcing
classical behavior in the macroscopic realm.  Decoherence results in
environment-induced superselection\cite{Zurek91,Zurek81,Zurek86b} which
destroys superpositions between the states of preferred pointer
basis\cite{Zurek81}.  Classical computers are already decohered --
computation takes them through a predictable sequence of such pointer
states, which are stable in spite of the environment.  Thus, classical
computers cannot be put in arbitrary superpositions and cannot take
advantage of the quantum factoring algorithm.  But coupling with the
environment will also be inevitable for any system employed to
implement the quantum factoring algorithm.  Here, we will show what
the effect of decoherence on the quantum factoring algorithm is.

Our model involves the introduction of the environment as a system external
to the computer.  Its state is represented by third label.
The input state may thus be written as
\begin{equation}
	|\tilde\psi_1\rangle = {1\over\sqrt{ q}}
		\sum_{a=0}^{q-1} |a,0\rangle\times|\epsilon\rangle
\,,
\end{equation}
where $\epsilon$ are the degrees of freedom of the environment.  The
environment is initially uncorrelated with the computer; however, it
is likely that the interaction between bits necessary for the
calculation of $x^a\ {\rm mod}\ N$ will involve some interaction with
the environment, so that the next state,
\begin{equation}
	|\tilde\psi_2\rangle = {1\over\sqrt{ q}}
		\sum_{a=0}^{q-1} |a, x^a {\rm mod}\ N\rangle
                \times|\epsilon_a\rangle
\,,
\end{equation}
leaves the environment partially correlated with the state of the
computer.  Now, if the physical representation for the computer's
quantum bits is diagonal in the pointer basis of the environment, then
decoherence results in no adverse effects when measuring the second
label of $|\tilde\psi_2\rangle$.  Such a design would be optimal.  We
thus focus on the effects of decoherence on the first label, by
suppressing the second label, and tracing over the environment to
obtain the reduced density matrix;
\begin{equation}
	\rho_{red} = {1\over q} \sum_{a=0}^{q-1}
		{\mbox{\hspace*{-2pt}\raisebox{3pt}{$\mathop{'}$}}}
		\sum_{a'=0}^{q-1}
		{\mbox{\hspace*{-2pt}\raisebox{3pt}{$\mathop{'}$}}}
		\,
		\left[\rule{0pt}{2.4ex}{ 1- \beta_{aa'} }\right] |a\rangle \langle a'|
\,.
\label{eq:rhored}
\end{equation}
Here $1-\beta_{aa'}= |\langle\epsilon_a|\epsilon_{a'}\rangle|^2$
is a measure of the accuracy with which the state of the environment
has become correlated with the state of the quantum computer.
If $|a\rangle$ and
$|a'\rangle$ are quantum bit register states diagonal in the pointer
basis, then we may take;
\begin{equation}
	1 - \beta_{aa'} \approx \exp \left[\rule{0pt}{2.4ex}{ -\xi (a\ \otimes\ a')
}\right] ,
\label{eq:betadef}
\end{equation}
where $\times$ is defined as the exclusive-or (XOR) function, and
gives the Hamming distance\cite{hamming} between $a$ and $a'$.  $\xi$
is a constant parameter which depends on the particular realization of
the quantum computer.  The measurement results in a probability
distribution, shown in Fig.~\ref{fig:decohered}, which differs from
the one in eq. (4) (see Fig. 1)
in that non-zero-probabilities have appeared between the peaks and
that these peaks have decreased in amplitude.

The qualitative effect of decoherence is well approximated by the
simpler function $1-\beta_{aa'}\equiv \delta_{aa'} +
(1-\delta_{aa'})\beta$, where $\beta$ is a constant.
For $\beta=0$ we get the state with complete
coherence and $\beta=1$ one with complete decoherence (i.e., a
matrix diagonal in the pointer state).  In the limit of $\beta\sim 1$,
we may understand
$\beta$ using the fractional amount
of {\em information lost to the environment}, expressed as $
S_f/S_{max} = 1-(1-\beta)^2$, where $S_{max}$ is the entropy of a completely
decohered computer and $S_f$ is the difference between
the entropy of the final state of
the computer and the one from the initial state (assumed to be nearly zero).
For $\beta\approx 0.5$ the probability between the
peaks (see Fig.~\ref{fig:decohered}) is equal to the one of the peaks,
and thus there is as much
chance to get a correct answer than a wrong one.  Once $(1-\beta)^{-1}
\sim {\cal O}(\exp [(\log N)^{1/3}])$, the quantum computer becomes as
inefficient as a classical one.  In this case it would take a number
of trials exponential in $(\log N)^{1/3}$ in order to factor the number $N$.

In principle it is easy to calculate $\beta$ from an experiment.  If we
allow the input to be in the superposition $|\uparrow_{in}\rangle +
|\downarrow_{in}\rangle$ then the coefficient $\beta$ is given by
\begin{equation}
	\beta = 1 - {  \rho_{\uparrow\downarrow}^{out}
               	+ \rho_{\downarrow\uparrow}^{out}
		\over
	       	\rho_{\uparrow\uparrow}^{out}
		+ \rho_{\downarrow\downarrow}^{out} }
\,.
\end{equation}
In a two slit experiment, it is the ratio between the amplitude of
the destructive and constructive interference on the screen also
called the fringe visibility function.

It is reasonable to assume, to first approximation, that the loss of
coherence is linear with the number of operations in the computation.
This is equivalent to saying that the environment keeps no memory of
the system as it evolves from one step to another.  Thus $1-\beta
\approx n_{op} \alpha$, where $\alpha$ is the coherence lost in a simple
operation (once coherence is lost in such systems it cannot be
regained).  When $\alpha$ is small it
can also be interpreted as the fractional information
loss per operation, i.e. such that $\Delta S/S_{max} = \alpha$.
$n_{op}$ is the operation count, $n_{op} \sim [\log N]^2$.  It is
therefore possible to estimate the total loss of quantum coherence by
studying only one part of the computer. To factor a number $N$ the
quantum algorithm uses ${\cal O}([\ln N]^2)$ operations and therefore
$\alpha^{-1} \sim {\cal O}([\ln N]^2)$.  The quantum computer allows
a significant amount of decoherence.  One of the reasons is that
factoring is in the class of the so-called NP functions, i.e.
functions which are hard to solve but once the answer is known it is
fairly easy to verify.  The quantum computer can therefore run until
we find the correct factors.

How much entropy is lost to the environment per step?  Designs for
quantum computers have been
suggested\cite{Lloyd93,Div94,Chuang94,Cirac94} and some possible
difficulties investigated\cite{Landauer94,Unruh94}.  Common to these
designs is the model of a simple two state system interacting with an
ensemble of oscillators (the environment), from which we can get an
idea for what $\alpha$ is.  We use as a Hamiltonian
\begin{equation}
	H={\Delta\over 2}\hat\sigma_x + \mu\hat\sigma_z\sum_n C_n q_n +
		\sum_n h_n
\,.
\end{equation}
where the $\sigma$'s are Pauli matrices, $q_n$ are the coordinates of
the environment oscillators and $h_n$ are harmonic oscillator
Hamiltonians (a cutoff $\Lambda$ is implicit).  It can be seen that
without the environment, a state of the system of the form
$|\uparrow\rangle$ turns into the state $(|\uparrow\rangle +
|\downarrow\rangle)/\sqrt{2}$ in a time $\pi/2\Delta$.  This a typical
step for a quantum computer.

Thus, the effect of a zero temperature environment is to decrease the
off-diagonal term of the density matrix by the amount\cite{Paz94}
\begin{equation}
	\alpha\sim {\mu^2 \eta \over 2\pi}
	    \left[{ - {\bf  C} -{\pi^2\over4}
		 + \log \left({{\Delta\over \Lambda}}\right) }\right]
\,.
\end{equation}
Here, ${\bf C}$ is Euler's constant and $\eta$ is the viscosity
coefficient determined by the spectral density of the oscillators.  A
non-zero temperature will further increase the value of $\alpha$.

With perfect operation, each execution trial gives a factor of $N$
with probability ${\cal O}(\log N)$, but with decoherence, the number
of trials required becomes ${\cal O}(\log N/(1-\beta))$.
 In terms of $\alpha$, we find that
with decoherence, the required number of executions of the quantum
algorithm to find a factor of $N$ is of order
\begin{equation}
	\mbox{Number of trials}
	  \sim \frac{L}{1-{L^2}\alpha}
\,,
\end{equation}
where $L = \log N$.  To give performance better than the classical
algorithm, we must therefore have that
\begin{equation}
	\frac{L}{1-{L^2}\alpha}
		\leq \exp \left[\rule{0pt}{2.4ex}{ L^{1/3} }\right]
\,.
\end{equation}

Using eq (12) and (14) we find that the largest number which can be
factored efficiently with a quantum computer is
\begin{equation}
	N \sim \exp \left[\rule{0pt}{2.4ex}{ \frac{1}{\sqrt{\alpha}} }\right]
	  \sim \exp \left[\rule{0pt}{2.4ex}{ \sqrt{ \frac{2\pi}{\mu^2\eta} } }\right]
\,.
\end{equation}
The quantum algorithm to factor numbers uses the quantum computer as a
huge interferometer.  When it is perfectly isolated, the interference
fringes give the important clue to the factors.  When isolation is
not perfect anymore there are chances that the result is irrelevant
for factoring.  The quantum computer is efficient as long as we can
discover the interference pattern in a number of trials less than the
one given by the classical algorithm.

Many are skeptical of the possibility of building useful quantum computers.
This attitude is largely fuelled by the inevitability of decoherence and the
fragility of the information encoded in coherent quantum
superposition\cite{Landauer94,Unruh94}.  In particular, error correction
which ensures reliability of classical computers may be very difficult to
accomplish quantum mechanically.  We note that these critical remarks,
while well taken,
are based on a classical paradigm:  Computers are useful when they always
(or almost always) get
right answers.   By contrast, as a result of decoherence
quantum computers will often give
wrong answers.  However, providing that the resulting probability
distribution still gives one a clue of what the right answers are,
quantum computers will be useful.
This is because for one-way functions (which form a substantial
class of NP-hard problems) verification is trivial.  Thus a probability
distribution emerging form a quantum computer may suffice:  Fringes in an
interference experiment allow one to identify the characteristic frequency
of a source even  when the contrast is far from perfect.

\proclaim Acknowledgment.

\noindent We would like to thank J.Anglin and J.P.Paz for useful conversations.





\begin{figure}[htbp]
%
\caption{Probability distribution for the measurement of $c$ in the
	state given in Eq.(\protect\ref{eq:shorstate}) with $N=21$,
	$q=128$, $x=5$, $k=3$.  The broadening of the peaks is a
	result of using discrete Fourier transform with $q$ possible
	modes; a continuous Fourier transform would have given delta
	functions.}
\label{fig:perfect}
\end{figure}

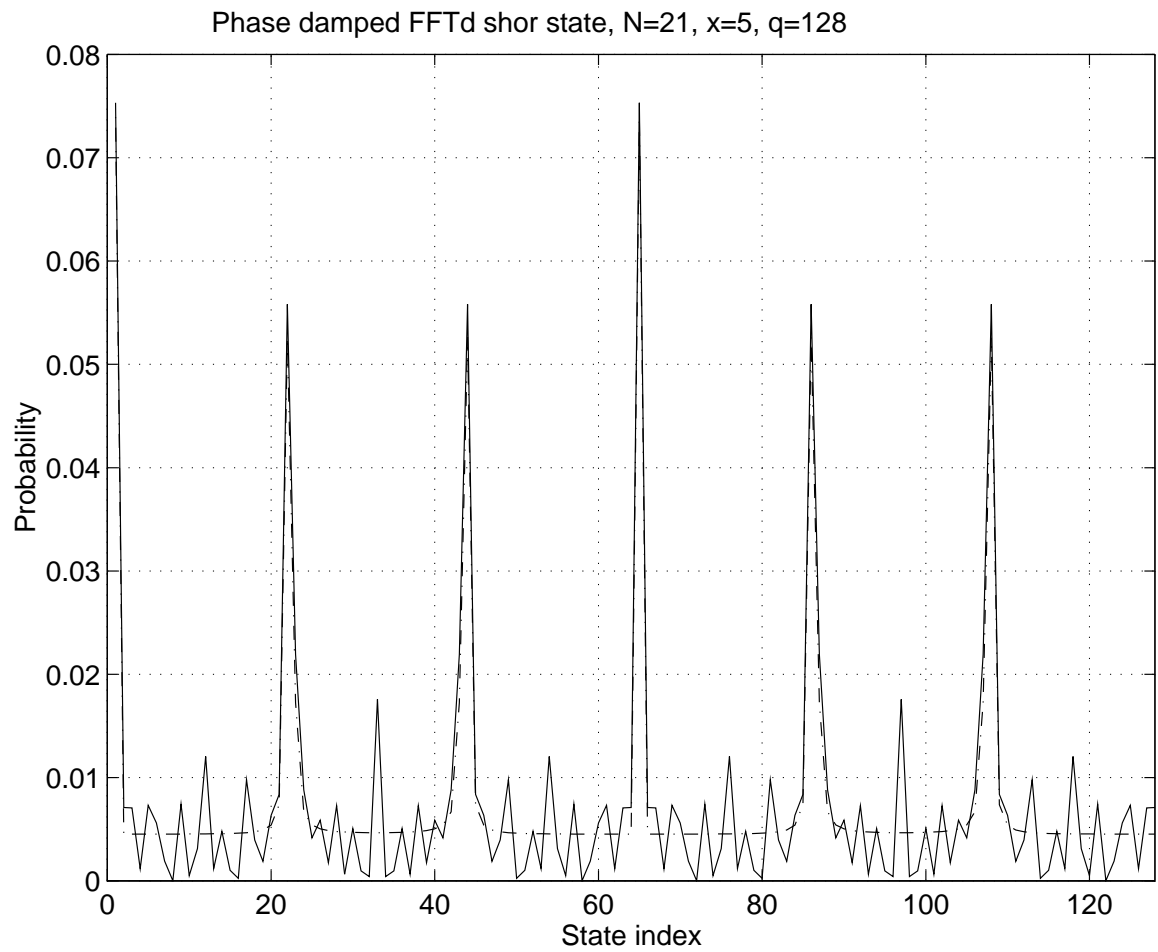
\begin{figure}[htbp]
%
\caption{Effect of decoherence on the probability distribution for the
	measurement of $c$. The state given by
	Eq.(\protect\ref{eq:rhored}) once $a$ if Fourier transformed.
	The decoherence parameter Eq.(\protect\ref{eq:betadef}) has
	been taken to be $\xi=0.1$.  The dotted line shows the good
	agreement of our constant-beta approximation, taking
	$\beta=0.58$.}
\label{fig:decohered}
\end{figure}

\end{document}